\documentclass[aps,prl,twocolumn,superscriptaddress,nofootinbib,nobibnotes,floatfix]{revtex4-2}
\usepackage{graphicx}
\usepackage[colorlinks=true,linkcolor=blue,citecolor=blue,urlcolor=blue]{hyperref}
\usepackage{amsmath,amssymb,bm}
\usepackage{xcolor}
\definecolor{osu}{HTML}{ba0c2f}
\usepackage{xurl}
\newcommand{\orcid}[1]{\texorpdfstring{\begingroup
  \hypersetup{hidelinks}\href{https://orcid.org/#1}{\includegraphics[width=10pt]{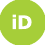}\,}%
\endgroup}{}}
\newcommand{\leff}{\ell_{\rm eff}}

\begin{document}

\title{Apollo 17 radio measurements test lunar subsurface models for Dark Ages 21-cm experiments}

\author{Paramita Dasgupta\,\orcid{0000-0002-9559-4803}}
\email{dasgupta.80@osu.edu}

\affiliation{Center for Cosmology and AstroParticle Physics (CCAPP), \href{https://ror.org/00rs6vg23}{\color{osu}The Ohio State University}, Columbus, OH 43210, USA}
\affiliation{Department of Physics, \href{https://ror.org/00rs6vg23}{\color{osu}The Ohio State University}, Columbus, OH 43210, USA}
\affiliation{Department of Astronomy, \href{https://ror.org/00rs6vg23}{\color{osu}The Ohio State University}, Columbus, OH 43210, USA}

\begin{abstract}
Below 50~MHz an antenna on the Moon couples to the regolith, so Dark Ages 21-cm experiments need subsurface models that predict its radio response. We test them with Apollo~17 surface-propagation data. A single buried interface fits 2.1 and 4.0~MHz best but predicts at 8.1--32.1~MHz an interference ripple up to 2.6 times the measured fluctuation. A graded profile predicts the data. All fits to 2.1--16~MHz reach permittivity 5 at 12--18~m depth. The data also disfavor the two-layer models of a LuSEE-Night study. A good fit in one band does not validate a subsurface model over the Dark Ages band.
\end{abstract}

\maketitle

In the standard cosmological model the Dark Ages, between recombination and the first stars, leave a sky-averaged 21-cm absorption signal of about 40~mK near 16~MHz, set by the thermal history of the gas and by any additional physics that changes it \cite{Furlanetto2006,PritchardLoeb2012,MondalBarkana2023}. The signal lies beneath Galactic emission more than $10^5$ times brighter, and below about 30~MHz the ionosphere and terrestrial interference hamper measurements from the Earth. This motivates radiometers on the lunar far side \cite{JesterFalcke2009,Koopmans2021,Burns2019}, such as LuSEE-Night \cite{Bale2023,BaleSlosar2024}. At these frequencies an antenna on the lunar surface or on a lander lies within a wavelength of the regolith, which becomes part of the antenna. The regolith sets the impedance, the beam and the fraction of the received power that comes from the ground, 36--88\% for a stand-alone LuSEE-Night-like dipole 1.5--3.5~m above a far-side reference profile \cite{DasguptaPRD}. At 90--250~K, the ground adds tens to hundreds of kelvin to the antenna temperature. Its frequency-dependent contribution must be modeled or fitted together with the 40~mK signal, with a subsurface model that describes how the real regolith reflects and emits. Refitting does not rescue a model of the wrong form. A two-layer model refitted to a far-side reference profile leaves a residual of about 4\% of the sky temperature, which overwhelms the signal \cite{DasguptaPRD}.

Design studies for these experiments have considered subsurface models ranging from a uniform half-space to stacks of two or eight uniform layers \cite{Burns2019,Yousuf2026,Camacho2026}, often represented by a surface impedance \cite{HFSSintro2025}. To our knowledge, none has been tested for its ability to predict radio measurements made on the Moon, although such measurements exist. In 1972, the Surface Electrical Properties (SEP) experiment of Apollo~17 measured radio propagation along the floor of the Taurus-Littrow valley at six frequencies from 1.0 to 32.1~MHz \cite{Simmons1972,Simmons1973}, across the band of these experiments. The SEP experiment team found that no plane-layer model reproduced all frequencies accurately \cite{SEPFinal1975}, and the archived measurements \cite{SEPdata2021} have since been inverted for a depth profile \cite{Grimm2018}. Here we ask a different question. Can a subsurface model selected by its fit in one band predict the measurements in another? With our layered-medium calculation, which includes the near field and every reflection in the regolith, we fit each model at 2.1 and 4.0~MHz and predict the measurements at 8.1, 16.0 and 32.1~MHz, recorded along the same traverse. Our comparison accounts for correlated residuals and the noise floor. We find that the best fit at 2.1 and 4.0~MHz, a single buried interface, predicts at the higher frequencies an interference ripple larger than the measured fluctuations, whereas a graded profile, whose permittivity rises with depth, predicts the measurements. A companion paper details the calculation and its verification and applies it to a far-side radiometer \cite{DasguptaPRD}.

\begin{figure}[t]
  \centering
  \includegraphics[width=\columnwidth]{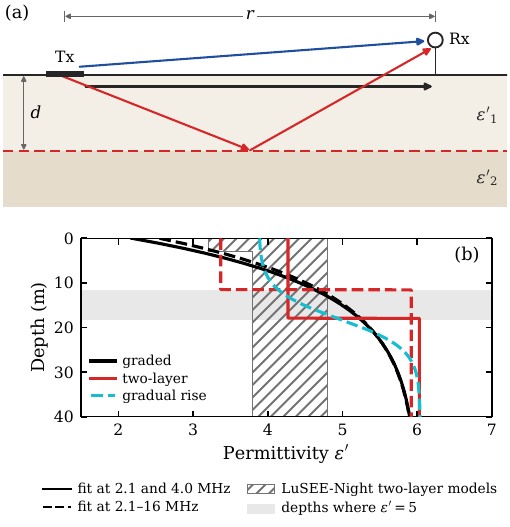}
  \caption{(a) The SEP experiment in side view, not to scale, with the transmitter (Tx) on the surface, the receiving loops (Rx) 2.5~m above it at range $r$, and the waves above the surface (blue), guided in the regolith (black) and reflected by a buried interface at depth $d$ (red). (b) Permittivity $\varepsilon'$ of the graded (black) and two-layer (red) profiles fitted to the Apollo~17 measurements at 2.1 and 4.0~MHz (solid) and at 2.1--16~MHz (dashed), and of the gradual transition profile fitted at 2.1--16~MHz (cyan, $\zeta=3.3$~m). Hatched: the two-layer models of a LuSEE-Night subsurface study \cite{Yousuf2026}. Shaded: the depths, 11.5--18.3~m, at which the fitted profiles reach $\varepsilon'=5$.}
  \label{fig:experiment}
\end{figure}

\textit{Measurements and calculation.}---Two orthogonal 70~m wires on the lunar surface formed the transmitter. Traps and suppressors made each wire a half-wave element at every frequency from 2.1 to 32.1~MHz \cite{Simmons1972}. Three orthogonal loops on the lunar rover, about 2.5~m above the surface, recorded the magnetic field as the astronauts drove away from the transmitter [Fig.~\ref{fig:experiment}(a)]. We decode the calibrated archive \cite{SEPdata2021} and keep the records taken during normal receiver operation, at transmitter--rover ranges of 50--1500~m. At each frequency $f$ we use three channels, in dB and binned in range (\hyperref[app:B]{Appendix~B}). These are the vertical field and the total horizontal power from the broadside transmitter, and the horizontal power from the endfire transmitter. We treat samples within 3~dB of the noise floor, estimated from the median level with the transmitter off, as upper limits. We leave out 1.0~MHz, at which the whole wire radiated with an undocumented current (\hyperref[app:B]{Appendix~B}).

The measured field shows an interference pattern \cite{Kong1972,Tsang1973}. The transmitter radiates a wave through the air and excites a wave guided in the regolith. The two travel at different phase velocities and beat along the traverse with a period of about $\lambda/(n-1)$, where $\lambda$ is the free-space wavelength and $n$ the refractive index of the regolith near the surface. A permittivity that increases with depth refracts the guided wave downward and damps the pattern, whereas a buried interface reflects part of the guided wave and adds a further ripple. We calculate the field over a planar stack of any number of lossy layers \cite{Dasgupta2026}. Layer $j$ has thickness $d_j$ and complex relative permittivity $\varepsilon_j=\varepsilon'_j(1+i\tan\delta)$, with one loss tangent $\tan\delta$ for all layers and time dependence $e^{-i\omega t}$, $\omega=2\pi f$. Layer $j=0$ is the vacuum above the surface and $j=N+1$ the substrate. The field of each current element is a superposition of plane waves of horizontal wavenumber $\kappa$ \cite{Sommerfeld1909,Weyl1919,Chew1995}, each reflected by the ground with the coefficient $\mathcal R_0$ that follows from the recursion \cite{Parratt1954}
\begin{equation}
  \mathcal R_{j-1}=\frac{r_{j-1,j}+\mathcal R_{j}\,e^{2i\xi_j}}{1+r_{j-1,j}\,\mathcal R_{j}\,e^{2i\xi_j}},
  \label{eq:parratt}
\end{equation}
started from $\mathcal R_N=r_{N,N+1}$ at the top of the substrate, for each polarization, $s$ and $p$. Here $r_{j-1,j}$ is the Fresnel reflection coefficient of the interface above layer $j$ (\hyperref[app:A]{Appendix~A}), $\xi_j=k_{z,j}d_j$ is the vertical phase across the layer, and $k_{z,j}=(\varepsilon_jk^2-\kappa^2)^{1/2}$, with ${\rm Im}\,k_{z,j}\ge0$ and $k=2\pi/\lambda$, is its vertical wavenumber. The reflected field, evanescent waves included, reduces to one-dimensional integrals over $\kappa$ (\hyperref[app:A]{Appendix~A}), which we integrate over the half-wave current of the transmitter. Our fields reproduce the image-dipole solution over a perfect conductor to a relative difference of $1.2\times10^{-7}$ and agree with the independent code \texttt{empymod} \cite{Werthmuller2017} for a half-space and two layered grounds to $1.4\times10^{-8}$ \cite{DasguptaPRD}.

We compare a lossy half-space, two uniform layers ($\varepsilon'_1$ over $\varepsilon'_2$ below a depth $d$) and three profiles in depth $z$. The graded profile, $\varepsilon'(z)=\varepsilon'_\infty-(\varepsilon'_\infty-\varepsilon'_0)e^{-z/z_g}$, rises from $\varepsilon'_0$ at the surface toward $\varepsilon'_\infty$ over a scale $z_g$. The transition profile, $\varepsilon'(z)=\varepsilon'_{\rm top}+(\varepsilon'_{\rm deep}-\varepsilon'_{\rm top})/[1+e^{-(z-z_t)/\zeta}]$, rises from $\varepsilon'_{\rm top}$ to $\varepsilon'_{\rm deep}$ around a depth $z_t$ over a scale $\zeta$, so it contains a single interface ($\zeta\to0$) as well as gradual rises. The third, the fixed surrogate $\varepsilon'(z)=6.1-2.4\,e^{-z/12\,{\rm m}}$ with $\tan\delta=0.012$, approximates the eight-layer profile with which a LuSEE-Night study \cite{Camacho2026} represents the inversion of the SEP data \cite{Grimm2018}. Each channel $c$ has a free offset $o_c$, for an instrumental constant in the archived levels. For the detected levels $D_{ci}$ of channel $c$ in range bin $i$, the predicted levels $D^{\rm mod}_{ci}$, the upper limits $D_c^{\rm up}$ of the censored bins, and an assumed scatter $\sigma_d=2$~dB per bin, we compare the descriptions with the score
\begin{multline}
  \leff=\sum_c\frac{N_{{\rm eff},c}}{N_c}\Big[\sum_{i\in{\rm det}}\ln\mathcal N(D_{ci}\,|\,D^{\rm mod}_{ci}+o_c,\sigma_d)\\
  +\sum_{i\in{\rm cens}}\ln\Phi\Big(\frac{D^{\rm up}_c-D^{\rm mod}_{ci}-o_c}{\sigma_d}\Big)\Big],
  \label{eq:elleff}
\end{multline}
where $\mathcal N$ and $\Phi$ are the normal density and cumulative distribution, and the sums run over detected and censored bins. The effective number of independent samples is $N_{{\rm eff},c}=N_c/\tau_c$, where $N_c$ is the number of bins and $\tau_c$ the integrated autocorrelation time of the residuals \cite{Geyer1992}, computed once and held fixed for all descriptions. Because the residuals are correlated, $\leff$ is a descriptive comparison score, not a likelihood, and we do not convert its differences into significance levels. Differences quoted are in $\leff$ unless stated otherwise. We test every comparison with an explicit covariance model, an Ornstein--Uhlenbeck (OU) process along the traverse for each channel, with five block-bootstrap schemes, and by removing the samples near the transmitter (\hyperref[app:B]{Appendix~B}).

\begin{figure*}[t]
  \centering
  \includegraphics[width=\textwidth]{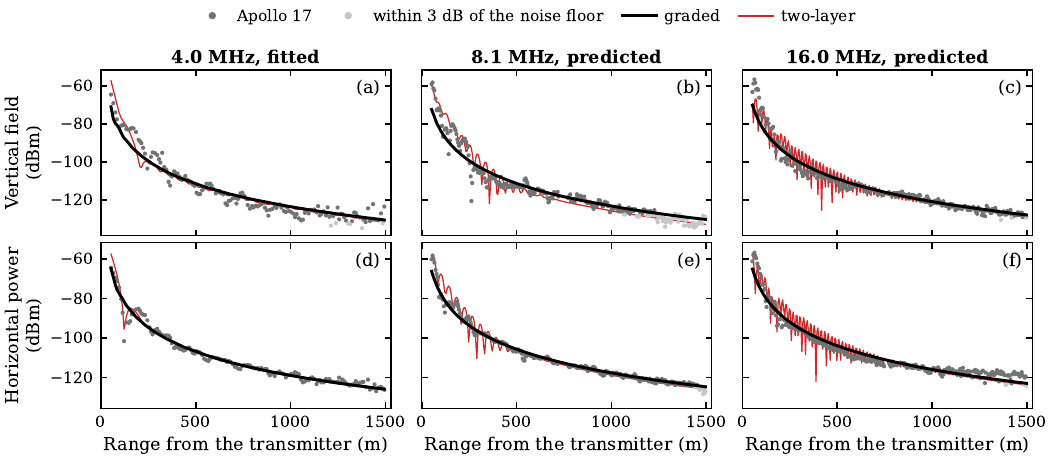}
  \caption{Apollo~17 broadside vertical field (top) and horizontal power (bottom) at 4.0~MHz, fitted jointly with 2.1~MHz, and at 8.1 and 16.0~MHz, predicted with the fitted parameters held fixed. Gray: measurements on the Moon. Light gray: samples within 3~dB of the noise floor. Black and red: the graded and two-layer descriptions fitted at 2.1 and 4.0~MHz, each shifted by its fitted channel offset. At 16.0~MHz, and at 8.1~MHz in the horizontal power, the two-layer description predicts a ripple larger than the measured fluctuations.}
  \label{fig:fields}
\end{figure*}

\textit{Fit at 2.1 and 4.0~MHz.}---The best fit is a two-layer regolith with a single buried interface, a 17.9~m layer with $\varepsilon'_1=4.27$ over a substrate with $\varepsilon'_2=6.03$, and $\tan\delta=0.008$ [Fig.~\ref{fig:experiment}(b), Table~\ref{tab:fits}]. It leads the best-fitting graded profile, which rises from $\varepsilon'_0=2.18$ to $\varepsilon'_\infty=6.01$ over a scale $z_g=11.3$~m, by 18.2 in $\leff$ and by 60.1 in the covariance-model log-likelihood. This preference persists in every analysis variant we tested, with differences of 8.4--31.8, and in 88--99\% of the bootstrap replicates for each scheme, each refitted in a linearized model \cite{DasguptaPRD}. With full refits where the linearization fails or the two descriptions nearly tie, the two-layer description is still favored in 76 of 80 replicates (\hyperref[app:B]{Appendix~B}). The lossy half-space scores 169.0 lower. Thus, using only the 2.1 and 4.0~MHz measurements, the two-layer description would be selected.

\begin{figure}[t]
  \centering
  \includegraphics[width=\columnwidth]{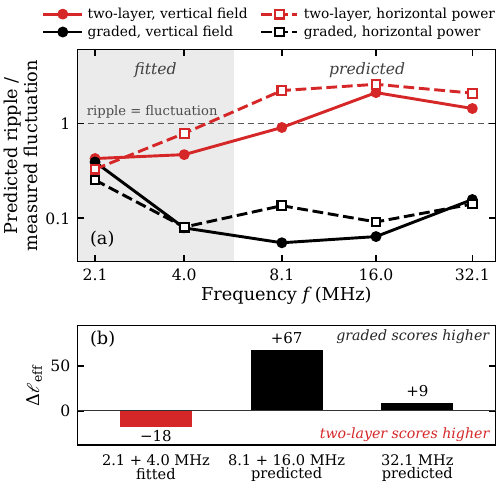}
  \caption{(a) Rms ripple predicted by the descriptions fitted at 2.1 and 4.0~MHz, divided by the measured rms fluctuation, for the two broadside channels (Table~\ref{tab:ripple}). The shaded band marks the fitted frequencies. Above the horizontal dashed line at 1, the predicted ripple exceeds the measured fluctuation. (b) Score of the graded profile minus that of the two-layer description, $\Delta\leff$, in each band (Table~\ref{tab:fits}). Black bars favor the graded profile and the red bar the two-layer description.}
  \label{fig:ripple}
\end{figure}

\textit{Prediction at 8.1--32.1~MHz.}---We now hold the subsurface parameters at their 2.1--4.0~MHz values, refit only the channel offsets and predict the higher frequencies [Figs.~\ref{fig:fields} and \ref{fig:ripple}(b), Table~\ref{tab:fits}]. The graded profile comes within 3.5 of the best-performing description, the fixed surrogate, at 8.1--16.0~MHz and within 0.1 at 32.1~MHz. The two-layer description scores 70.9 below the best at 8.1--16.0~MHz, 20 times the deficit of the graded profile, and 9.2 below at 32.1~MHz. Its deficits in the covariance-model log-likelihood are 788.9 and 631.0. Every description without a buried interface, including the lossy half-space, comes within 5.7 of the best. The two-layer description fails because its buried interface reflects part of the guided wave, which interferes with the other components and produces a ripple along the traverse, with a period of about 19~m at 16.0~MHz [Figs.~\ref{fig:fields}(c) and \ref{fig:fields}(f)].

The failure is also visible directly in the data. We subtract from each binned series a smooth trend, its Gaussian smoothing with a standard deviation of 25~m in range, and calculate the rms of the residuals over 100--800~m [Fig.~\ref{fig:ripple}(a), Table~\ref{tab:ripple}]. At 16.0~MHz, the two-layer description predicts ripples of 4.0 and 3.7~dB rms in the two broadside channels, whereas the measured fluctuations, rock scattering and noise included, are only 1.9 and 1.4~dB rms. If the predicted ripple were present and independent of the other fluctuations, their variances would add, so the measured rms would be expected to be at least the ripple rms. The predicted ripple also exceeds the measured fluctuation in the horizontal power at 8.1~MHz and in both broadside channels at 32.1~MHz, by factors of 1.4--2.2. These ratios persist for other trend widths and range windows (Table~\ref{tab:ripple}). The only exception is the vertical field at 8.1~MHz, 2.8 against 3.1~dB rms. Shifts, scale errors and even independent errors of 6~m rms in the archived transmitter--rover ranges \cite{SEPdata2021} leave the predicted ripple above the measured fluctuation in the same five channels \cite{DasguptaPRD}. At 2.1 and 4.0~MHz, by contrast, the measured fluctuations of 2.0--2.9~dB rms exceed the ripples predicted by both descriptions and mask them, so the low-frequency fit cannot reject the buried interface. The graded profile predicts ripples of at most 0.2~dB rms over 4.0--32.1~MHz.

We find the result robust to resampling and alternative analysis choices \cite{DasguptaPRD}. With the subsurface parameters fixed, block resampling favors the graded profile over the two-layer description at 8.1--16.0~MHz in 96--100\% of 500 replicates under each of five schemes, and at 32.1~MHz in every replicate. When we refit all descriptions without the samples within 100, 200 or 300~m of the transmitter, the two-layer description stays 90--201 below the graded profile at 8.1--16.0~MHz. The deficit is 65.4--68.6 for nine choices of transmitter current, including one solved by the method of moments over each profile, and 29--247 under alternative choices of receiver height, scatter, noise floor and autocorrelation times \cite{DasguptaPRD}. The second maximum of the two-layer fit, 24.1 below the best, also trails the graded profile at 8.1--16.0~MHz, by 25.3 (Table~\ref{tab:fits}). Its predicted ripple at 16.0~MHz likewise exceeds the measured fluctuation (Table~\ref{tab:ripple}).

The fitted loss tangents are low, and greater loss weakens the reflection responsible for the ripple. We therefore refit both descriptions with the loss tangent fixed in turn at 0.012, 0.020, 0.028 and 0.040. These values bracket the laboratory expectation of 0.016--0.026 for the mean composition of soils at the site \cite{Olhoeft1975,RobinsonJolliff2002}. The 2.1--4.0~MHz measurements continue to favor an interface near 18~m, which then scores 12--49 below the graded profile at 8.1--16.0~MHz \cite{DasguptaPRD}. Its predicted ripple at 16.0~MHz exceeds the measured fluctuation in at least one broadside channel for $\tan\delta\leq0.028$, but not at 0.040.

\textit{What the measurements constrain.}---To test how sharp the rise is, we fit the transition profile to 2.1--16~MHz jointly. The best fit is a gradual rise from $\varepsilon'_{\rm top}=3.88$ to $\varepsilon'_{\rm deep}=6.04$, centered at $z_t=18.1$~m, with $\zeta=3.3$~m (Table~\ref{tab:joint}). A sharp rise, with $\zeta=0.05$--0.5~m, scores only 3.8--4.4 lower \cite{DasguptaPRD}. It coincides with the two-layer description fitted to both bands, an 11.5~m layer with $\varepsilon'_1=3.37$ over $\varepsilon'_2=5.92$, which fits better than the graded profile by 11.4 (Table~\ref{tab:joint}). The two bands favor different profiles, so the sharpness is not determined. At 2.1--4.0~MHz the gradual rise leads by 21.1. At 8.1--16.0~MHz the sharp rise leads by 17.2, and 16.9 of this comes from one channel, the broadside vertical field within 300~m of the transmitter, which is the most sensitive to rocks and lateral variations \cite{DasguptaPRD}. A rough interface acts like a gradual rise, because for the coherent field an interface whose height has a Gaussian distribution of rms $\sigma_h$ reflects like an error-function transition of width $\sigma_h$ \cite{NevotCroce1980}. The scale $\zeta$ thus measures the vertical extent of the transition, whether from a graded subsurface or a rough interface.

The depth is much better constrained. In every fit to 2.1--16~MHz, the permittivity reaches 5 at a depth $z_5=11.5$--18.3~m, has a mean $\langle\varepsilon'\rangle=3.4$--4.2 above that depth and is 5.9--6.0 at 100~m [Fig.~\ref{fig:experiment}(b), Table~\ref{tab:joint}]. The measurements therefore require a substantial rise of the permittivity with depth at Taurus-Littrow. Through the laboratory relation $\varepsilon'=1.919^\rho$, with the bulk density $\rho$ in g\,cm$^{-3}$ \cite{Olhoeft1975}, the permittivity implies densities of 1.9--2.2~g\,cm$^{-3}$ above the rise and 2.7~g\,cm$^{-3}$, or about 20\% porosity, at 100~m. The rise lies below the base of the regolith, 4--8~m deep at the site \cite{Cooper1974,Bugiolacchi2016}, within fractured material identified by seismic profiling \cite{Cooper1974,Heffels2017}. Within the adopted relation, this material therefore becomes denser with depth, as the earlier inversion of the SEP data also found \cite{Grimm2018}. The failure is that of the two-layer profile selected at 2.1--4.0~MHz, not of every two-layer structure. The reverse test, fitting 8.1--16.0~MHz and predicting 2.1--4.0~MHz, confirms that the low frequencies fix the deep permittivity (\hyperref[app:B]{Appendix~B}).

\textit{Implications.}---These results have three implications for 21-cm experiments. First, the two-layer models of a LuSEE-Night subsurface study, $\varepsilon'=3.2$--4.2 over 3.8--4.8 below 0.5--3~m \cite{Yousuf2026}, already fail to fit the 2.1--4.0~MHz measurements at Taurus-Littrow. Every model in a grid spanning this range scores at least 328 below the graded profile, more than twice the 150.8 by which a uniform half-space trails it (Table~\ref{tab:fits}). With the loss tangent free, the best still trails by 111 \cite{DasguptaPRD}. These frequencies probe tens of meters, where the measurements require $\varepsilon'\approx6$ but the models reach only 4.8. Second, current LuSEE-Night simulations \cite{Yousuf2026,Camacho2026} replace the layered regolith by a surface impedance, which is exact only at normal incidence \cite{HFSSintro2025,SeniorVolakis1995}. Our calculation makes no such approximation. Refitted with the surface impedance, the graded profile fits 2.1--4.0~MHz worse by 11.5, and its permittivity rises only from 2.33 to 2.6 at 50~m, instead of to 6.0 \cite{DasguptaPRD}. For a stand-alone LuSEE-Night-like dipole 0.3--3.5~m above a far-side reference profile, the surface impedance misestimates the fraction of received power contributed by the ground by 13--17 percentage points \cite{DasguptaPRD}. Third, the sharpness of the transition, which the measurements leave open, matters for arrays on the surface. Across 5--45~MHz the zenith directivity of a dipole lying on the regolith varies by 1--2\% for the gradual rises fitted to 2.1--16~MHz, but by a factor of 1.9 for the sharp ones \cite{DasguptaPRD}.

These results apply to the Apollo~17 near-side site. No equivalent measurements exist on the far side, where subsurface information comes from radar soundings at much higher frequencies \cite{Li2020,Zhang2025CE6}. A far-side subsurface model should therefore allow the permittivity to rise over the top tens of meters, with both the depth and the sharpness of the rise left free. It should be tested against measurements at frequencies withheld from its fit, such as passes of an orbiting calibrator over the landing site \cite{Bale2023}.

A subsurface model that fits lunar radio measurements in one band need not predict them in another. At Taurus-Littrow, the interface preferred at 2.1--4.0~MHz predicts at 8.1--32.1~MHz a ripple larger than the measured fluctuations in five of six broadside channels, whereas a permittivity rising with depth predicts the measurements across a factor of 15 in frequency. Because the Apollo~17 measurements are public, any subsurface model or simulation code can be tested in this way against radio data from the Moon. Lunar 21-cm experiments should therefore select subsurface models by their predictive performance, not by their fit in one band. The same principle applies on Earth, where the ground beneath an antenna can also shape the measured spectrum \cite{Bradley2019}.

\textit{Data availability.}---The Apollo~17 SEP measurements are available from the NASA Planetary Data System \cite{SEPdata2021}.

\begin{acknowledgments}
The author acknowledges the Apollo~17 crew and the Surface Electrical Properties (SEP) experiment team for acquiring and documenting the lunar measurements analyzed in this work, and the NASA Planetary Data System for restoring, preserving and providing access to the archived SEP data and associated documentation. This work was supported by the Center for Cosmology and AstroParticle Physics (CCAPP) at The Ohio State University.
\end{acknowledgments}

\bibliography{lunar_prl}

\begin{thebibliography}{38}%
\makeatletter
\providecommand \@ifxundefined [1]{%
 \@ifx{#1\undefined}
}%
\providecommand \@ifnum [1]{%
 \ifnum #1\expandafter \@firstoftwo
 \else \expandafter \@secondoftwo
 \fi
}%
\providecommand \@ifx [1]{%
 \ifx #1\expandafter \@firstoftwo
 \else \expandafter \@secondoftwo
 \fi
}%
\providecommand \natexlab [1]{#1}%
\providecommand \enquote  [1]{``#1''}%
\providecommand \bibnamefont  [1]{#1}%
\providecommand \bibfnamefont [1]{#1}%
\providecommand \citenamefont [1]{#1}%
\providecommand \href@noop [0]{\@secondoftwo}%
\providecommand \href [0]{\begingroup \@sanitize@url \@href}%
\providecommand \@href[1]{\@@startlink{#1}\@@href}%
\providecommand \@@href[1]{\endgroup#1\@@endlink}%
\providecommand \@sanitize@url [0]{\catcode `\\12\catcode `\$12\catcode
  `\&12\catcode `\#12\catcode `\^12\catcode `\_12\catcode `\%12\relax}%
\providecommand \@@startlink[1]{}%
\providecommand \@@endlink[0]{}%
\providecommand \url  [0]{\begingroup\@sanitize@url \@url }%
\providecommand \@url [1]{\endgroup\@href {#1}{\urlprefix }}%
\providecommand \urlprefix  [0]{URL }%
\providecommand \Eprint [0]{\href }%
\providecommand \doibase [0]{https://doi.org/}%
\providecommand \selectlanguage [0]{\@gobble}%
\providecommand \bibinfo  [0]{\@secondoftwo}%
\providecommand \bibfield  [0]{\@secondoftwo}%
\providecommand \translation [1]{[#1]}%
\providecommand \BibitemOpen [0]{}%
\providecommand \bibitemStop [0]{}%
\providecommand \bibitemNoStop [0]{.\EOS\space}%
\providecommand \EOS [0]{\spacefactor3000\relax}%
\providecommand \BibitemShut  [1]{\csname bibitem#1\endcsname}%
\let\auto@bib@innerbib\@empty
\bibitem [{\citenamefont {Furlanetto}\ \emph {et~al.}(2006)\citenamefont
  {Furlanetto}, \citenamefont {Oh},\ and\ \citenamefont
  {Briggs}}]{Furlanetto2006}%
  \BibitemOpen
  \bibfield  {author} {\bibinfo {author} {\bibfnamefont {S.~R.}\ \bibnamefont
  {Furlanetto}}, \bibinfo {author} {\bibfnamefont {S.~P.}\ \bibnamefont {Oh}},\
  and\ \bibinfo {author} {\bibfnamefont {F.~H.}\ \bibnamefont {Briggs}},\
  }\bibfield  {title} {\bibinfo {title} {Cosmology at low frequencies: The 21
  cm transition and the high-redshift {U}niverse},\ }\href
  {https://doi.org/10.1016/j.physrep.2006.08.002} {\bibfield  {journal}
  {\bibinfo  {journal} {Phys. Rep.}\ }\textbf {\bibinfo {volume} {433}},\
  \bibinfo {pages} {181} (\bibinfo {year} {2006})},\ \Eprint
  {https://arxiv.org/abs/astro-ph/0608032} {arXiv:astro-ph/0608032}
  \BibitemShut {NoStop}%
\bibitem [{\citenamefont {Pritchard}\ and\ \citenamefont
  {Loeb}(2012)}]{PritchardLoeb2012}%
  \BibitemOpen
  \bibfield  {author} {\bibinfo {author} {\bibfnamefont {J.~R.}\ \bibnamefont
  {Pritchard}}\ and\ \bibinfo {author} {\bibfnamefont {A.}~\bibnamefont
  {Loeb}},\ }\bibfield  {title} {\bibinfo {title} {21 cm cosmology in the 21st
  century},\ }\href {https://doi.org/10.1088/0034-4885/75/8/086901} {\bibfield
  {journal} {\bibinfo  {journal} {Rep. Prog. Phys.}\ }\textbf {\bibinfo
  {volume} {75}},\ \bibinfo {pages} {086901} (\bibinfo {year} {2012})},\
  \Eprint {https://arxiv.org/abs/1109.6012} {arXiv:1109.6012} \BibitemShut
  {NoStop}%
\bibitem [{\citenamefont {Mondal}\ and\ \citenamefont
  {Barkana}(2023)}]{MondalBarkana2023}%
  \BibitemOpen
  \bibfield  {author} {\bibinfo {author} {\bibfnamefont {R.}~\bibnamefont
  {Mondal}}\ and\ \bibinfo {author} {\bibfnamefont {R.}~\bibnamefont
  {Barkana}},\ }\bibfield  {title} {\bibinfo {title} {Prospects for precision
  cosmology with the 21 cm signal from the dark ages},\ }\href
  {https://doi.org/10.1038/s41550-023-02057-y} {\bibfield  {journal} {\bibinfo
  {journal} {Nat. Astron.}\ }\textbf {\bibinfo {volume} {7}},\ \bibinfo {pages}
  {1025} (\bibinfo {year} {2023})},\ \Eprint {https://arxiv.org/abs/2305.08593}
  {arXiv:2305.08593} \BibitemShut {NoStop}%
\bibitem [{\citenamefont {Jester}\ and\ \citenamefont
  {Falcke}(2009)}]{JesterFalcke2009}%
  \BibitemOpen
  \bibfield  {author} {\bibinfo {author} {\bibfnamefont {S.}~\bibnamefont
  {Jester}}\ and\ \bibinfo {author} {\bibfnamefont {H.}~\bibnamefont
  {Falcke}},\ }\bibfield  {title} {\bibinfo {title} {Science with a lunar
  low-frequency array: From the dark ages of the {U}niverse to nearby
  exoplanets},\ }\href {https://doi.org/10.1016/j.newar.2009.02.001} {\bibfield
   {journal} {\bibinfo  {journal} {New Astron. Rev.}\ }\textbf {\bibinfo
  {volume} {53}},\ \bibinfo {pages} {1} (\bibinfo {year} {2009})},\ \Eprint
  {https://arxiv.org/abs/0902.0493} {arXiv:0902.0493} \BibitemShut {NoStop}%
\bibitem [{\citenamefont {Koopmans}\ \emph {et~al.}(2021)\citenamefont
  {Koopmans}, \citenamefont {Barkana}, \citenamefont {Bentum}, \citenamefont
  {Bernardi}, \citenamefont {Boonstra} \emph {et~al.}}]{Koopmans2021}%
  \BibitemOpen
  \bibfield  {author} {\bibinfo {author} {\bibfnamefont {L.~V.~E.}\
  \bibnamefont {Koopmans}}, \bibinfo {author} {\bibfnamefont {R.}~\bibnamefont
  {Barkana}}, \bibinfo {author} {\bibfnamefont {M.}~\bibnamefont {Bentum}},
  \bibinfo {author} {\bibfnamefont {G.}~\bibnamefont {Bernardi}}, \bibinfo
  {author} {\bibfnamefont {A.-J.}\ \bibnamefont {Boonstra}}, \emph {et~al.},\
  }\bibfield  {title} {\bibinfo {title} {Peering into the dark (ages) with
  low-frequency space interferometers: Using the 21-cm signal of neutral
  hydrogen from the infant universe to probe fundamental (astro)physics},\
  }\href {https://doi.org/10.1007/s10686-021-09743-7} {\bibfield  {journal}
  {\bibinfo  {journal} {Exp. Astron.}\ }\textbf {\bibinfo {volume} {51}},\
  \bibinfo {pages} {1641} (\bibinfo {year} {2021})}\BibitemShut {NoStop}%
\bibitem [{\citenamefont {Burns}\ \emph {et~al.}(2019)\citenamefont {Burns},
  \citenamefont {Hallinan}, \citenamefont {Lux} \emph {et~al.}}]{Burns2019}%
  \BibitemOpen
  \bibfield  {author} {\bibinfo {author} {\bibfnamefont {J.~O.}\ \bibnamefont
  {Burns}}, \bibinfo {author} {\bibfnamefont {G.}~\bibnamefont {Hallinan}},
  \bibinfo {author} {\bibfnamefont {J.}~\bibnamefont {Lux}}, \emph {et~al.},\
  }\href@noop {} {\bibinfo {title} {{NASA} probe study report: {F}arside
  {A}rray for {R}adio {S}cience {I}nvestigations of the {D}ark ages and
  {E}xoplanets ({FARSIDE})}} (\bibinfo {year} {2019}),\ \Eprint
  {https://arxiv.org/abs/1911.08649} {arXiv:1911.08649} \BibitemShut {NoStop}%
\bibitem [{\citenamefont {Bale}\ \emph {et~al.}(2023)\citenamefont {Bale} \emph
  {et~al.}}]{Bale2023}%
  \BibitemOpen
  \bibfield  {author} {\bibinfo {author} {\bibfnamefont {S.~D.}\ \bibnamefont
  {Bale}} \emph {et~al.},\ }\href@noop {} {\bibinfo {title} {{LuSEE} `{N}ight':
  The {L}unar {S}urface {E}lectromagnetics {E}xperiment}} (\bibinfo {year}
  {2023}),\ \Eprint {https://arxiv.org/abs/2301.10345} {arXiv:2301.10345}
  \BibitemShut {NoStop}%
\bibitem [{\citenamefont {Bale}\ and\ \citenamefont
  {Slosar}(2024)}]{BaleSlosar2024}%
  \BibitemOpen
  \bibfield  {author} {\bibinfo {author} {\bibfnamefont {S.~D.}\ \bibnamefont
  {Bale}}\ and\ \bibinfo {author} {\bibfnamefont {A.}~\bibnamefont {Slosar}},\
  }\bibfield  {title} {\bibinfo {title} {{LuSEE Night}: the lunar surface
  electromagnetics experiment},\ }in\ \href
  {https://doi.org/10.1117/12.3019515} {\emph {\bibinfo {booktitle} {Space
  Telescopes and Instrumentation 2024: Optical, Infrared, and Millimeter
  Wave}}},\ \bibinfo {series} {Proc. SPIE}, Vol.\ \bibinfo {volume} {13092}\
  (\bibinfo {year} {2024})\ p.\ \bibinfo {pages} {130922O}\BibitemShut
  {NoStop}%
\bibitem [{\citenamefont {Dasgupta}(2026{\natexlab{a}})}]{DasguptaPRD}%
  \BibitemOpen
  \bibfield  {author} {\bibinfo {author} {\bibfnamefont {P.}~\bibnamefont
  {Dasgupta}},\ }\href@noop {} {\bibinfo {title} {Lunar regolith effects on
  low-frequency antennas for {Dark Ages} 21-cm measurements: a self-consistent
  layered-medium model tested against {Apollo}~17 data}} (\bibinfo {year}
  {2026}{\natexlab{a}}),\ \bibinfo {note} {companion paper}\BibitemShut
  {NoStop}%
\bibitem [{\citenamefont {Yousuf}\ \emph {et~al.}(2026)\citenamefont {Yousuf},
  \citenamefont {Li}, \citenamefont {Bale}, \citenamefont {Barker},
  \citenamefont {Burns} \emph {et~al.}}]{Yousuf2026}%
  \BibitemOpen
  \bibfield  {author} {\bibinfo {author} {\bibfnamefont {F.}~\bibnamefont
  {Yousuf}}, \bibinfo {author} {\bibfnamefont {Z.}~\bibnamefont {Li}}, \bibinfo
  {author} {\bibfnamefont {S.~D.}\ \bibnamefont {Bale}}, \bibinfo {author}
  {\bibfnamefont {D.~W.}\ \bibnamefont {Barker}}, \bibinfo {author}
  {\bibfnamefont {J.~O.}\ \bibnamefont {Burns}}, \emph {et~al.},\ }\href@noop
  {} {\bibinfo {title} {Joint estimation of properties of the lunar subsurface
  and {G}alactic foregrounds with {LuSEE-Night}}} (\bibinfo {year} {2026}),\
  \bibinfo {note} {{LuSEE-Night} Collaboration},\ \Eprint
  {https://arxiv.org/abs/2604.21170} {arXiv:2604.21170} \BibitemShut {NoStop}%
\bibitem [{\citenamefont {Camacho}\ \emph {et~al.}(2026)\citenamefont
  {Camacho}, \citenamefont {Rotermund}, \citenamefont {Slosar}, \citenamefont
  {Bale}, \citenamefont {Barker} \emph {et~al.}}]{Camacho2026}%
  \BibitemOpen
  \bibfield  {author} {\bibinfo {author} {\bibfnamefont {H.}~\bibnamefont
  {Camacho}}, \bibinfo {author} {\bibfnamefont {K.~M.}\ \bibnamefont
  {Rotermund}}, \bibinfo {author} {\bibfnamefont {A.}~\bibnamefont {Slosar}},
  \bibinfo {author} {\bibfnamefont {S.~D.}\ \bibnamefont {Bale}}, \bibinfo
  {author} {\bibfnamefont {D.~W.}\ \bibnamefont {Barker}}, \emph {et~al.},\
  }\bibfield  {title} {\bibinfo {title} {Linear map-making with
  {LuSEE-Night}},\ }\bibfield  {journal} {\bibinfo  {journal} {Open J.
  Astrophys.}\ }\textbf {\bibinfo {volume} {9}},\ \href
  {https://doi.org/10.33232/001c.158626} {10.33232/001c.158626} (\bibinfo
  {year} {2026}),\ \Eprint {https://arxiv.org/abs/2508.16773}
  {arXiv:2508.16773} \BibitemShut {NoStop}%
\bibitem [{\citenamefont {{ANSYS, Inc.}}(2025)}]{HFSSintro2025}%
  \BibitemOpen
  \bibfield  {author} {\bibinfo {author} {\bibnamefont {{ANSYS, Inc.}}},\
  }\href@noop {} {\emph {\bibinfo {title} {An Introduction to {HFSS}}}}
  (\bibinfo {year} {2025}),\ \bibinfo {note} {release 2025 R2, Sec.~2, Layered
  Impedance}\BibitemShut {NoStop}%
\bibitem [{\citenamefont {Simmons}\ \emph {et~al.}(1972)\citenamefont
  {Simmons}, \citenamefont {Strangway}, \citenamefont {Bannister},
  \citenamefont {Baker}, \citenamefont {Cubley}, \citenamefont {LaTorraca},\
  and\ \citenamefont {Watts}}]{Simmons1972}%
  \BibitemOpen
  \bibfield  {author} {\bibinfo {author} {\bibfnamefont {G.}~\bibnamefont
  {Simmons}}, \bibinfo {author} {\bibfnamefont {D.~W.}\ \bibnamefont
  {Strangway}}, \bibinfo {author} {\bibfnamefont {L.}~\bibnamefont
  {Bannister}}, \bibinfo {author} {\bibfnamefont {R.}~\bibnamefont {Baker}},
  \bibinfo {author} {\bibfnamefont {D.}~\bibnamefont {Cubley}}, \bibinfo
  {author} {\bibfnamefont {G.}~\bibnamefont {LaTorraca}},\ and\ \bibinfo
  {author} {\bibfnamefont {R.}~\bibnamefont {Watts}},\ }\bibfield  {title}
  {\bibinfo {title} {The surface electrical properties experiment},\ }in\
  \href@noop {} {\emph {\bibinfo {booktitle} {Lunar Geophysics}}},\ \bibinfo
  {editor} {edited by\ \bibinfo {editor} {\bibfnamefont {Z.}~\bibnamefont
  {Kopal}}\ and\ \bibinfo {editor} {\bibfnamefont {D.~W.}\ \bibnamefont
  {Strangway}}}\ (\bibinfo  {publisher} {D. Reidel},\ \bibinfo {address}
  {Dordrecht},\ \bibinfo {year} {1972})\ pp.\ \bibinfo {pages}
  {258--271}\BibitemShut {NoStop}%
\bibitem [{\citenamefont {Simmons}\ \emph {et~al.}(1973)\citenamefont
  {Simmons}, \citenamefont {Strangway}, \citenamefont {Annan}, \citenamefont
  {Baker}, \citenamefont {Bannister}, \citenamefont {Brown}, \citenamefont
  {Cooper}, \citenamefont {Cubley}, \citenamefont {deBettencourt},
  \citenamefont {England}, \citenamefont {Groener}, \citenamefont {Kong},
  \citenamefont {LaTorraca}, \citenamefont {Meyer}, \citenamefont {Nanda},
  \citenamefont {Redman}, \citenamefont {Rossiter}, \citenamefont {Tsang},
  \citenamefont {Urner},\ and\ \citenamefont {Watts}}]{Simmons1973}%
  \BibitemOpen
  \bibfield  {author} {\bibinfo {author} {\bibfnamefont {G.}~\bibnamefont
  {Simmons}}, \bibinfo {author} {\bibfnamefont {D.}~\bibnamefont {Strangway}},
  \bibinfo {author} {\bibfnamefont {P.}~\bibnamefont {Annan}}, \bibinfo
  {author} {\bibfnamefont {R.~G.}\ \bibnamefont {Baker}}, \bibinfo {author}
  {\bibfnamefont {L.}~\bibnamefont {Bannister}}, \bibinfo {author}
  {\bibfnamefont {R.~E.}\ \bibnamefont {Brown}}, \bibinfo {author}
  {\bibfnamefont {W.}~\bibnamefont {Cooper}}, \bibinfo {author} {\bibfnamefont
  {D.}~\bibnamefont {Cubley}}, \bibinfo {author} {\bibfnamefont
  {J.}~\bibnamefont {deBettencourt}}, \bibinfo {author} {\bibfnamefont {A.~W.}\
  \bibnamefont {England}}, \bibinfo {author} {\bibfnamefont {J.}~\bibnamefont
  {Groener}}, \bibinfo {author} {\bibfnamefont {J.}~\bibnamefont {Kong}},
  \bibinfo {author} {\bibfnamefont {G.}~\bibnamefont {LaTorraca}}, \bibinfo
  {author} {\bibfnamefont {J.}~\bibnamefont {Meyer}}, \bibinfo {author}
  {\bibfnamefont {V.}~\bibnamefont {Nanda}}, \bibinfo {author} {\bibfnamefont
  {D.}~\bibnamefont {Redman}}, \bibinfo {author} {\bibfnamefont
  {J.}~\bibnamefont {Rossiter}}, \bibinfo {author} {\bibfnamefont
  {L.}~\bibnamefont {Tsang}}, \bibinfo {author} {\bibfnamefont
  {J.}~\bibnamefont {Urner}},\ and\ \bibinfo {author} {\bibfnamefont
  {R.}~\bibnamefont {Watts}},\ }\bibfield  {title} {\bibinfo {title} {Surface
  electrical properties experiment},\ }in\ \href
  {https://ntrs.nasa.gov/citations/19740010332} {\emph {\bibinfo {booktitle}
  {Apollo 17 Preliminary Science Report}}}\ (\bibinfo  {publisher} {NASA},\
  \bibinfo {address} {Washington, DC},\ \bibinfo {year} {1973})\ \bibinfo
  {note} {{NASA} SP-330, Sec.~15}\BibitemShut {NoStop}%
\bibitem [{\citenamefont {Simmons}(1974)}]{SEPFinal1975}%
  \BibitemOpen
  \bibfield  {author} {\bibinfo {author} {\bibfnamefont {G.}~\bibnamefont
  {Simmons}},\ }\href {https://ntrs.nasa.gov/citations/19750013120} {\emph
  {\bibinfo {title} {Surface electrical properties experiment}}},\ \bibinfo
  {type} {Technical Report}\ \bibinfo {number} {CSR-TR-74-1}\ (\bibinfo
  {institution} {Center for Space Research, Massachusetts Institute of
  Technology},\ \bibinfo {year} {1974})\ \bibinfo {note}
  {{NASA-CR-141724}}\BibitemShut {NoStop}%
\bibitem [{\citenamefont {{NASA Planetary Data System}}(2021)}]{SEPdata2021}%
  \BibitemOpen
  \bibfield  {author} {\bibinfo {author} {\bibnamefont {{NASA Planetary Data
  System}}},\ }\href {https://doi.org/10.17189/1518359} {\bibinfo {title}
  {Apollo 17 surface electrical properties experiment calibrated {ASCII} data
  bundle}} (\bibinfo {year} {2021}),\ \bibinfo {note} {urn:nasa:pds:a17sep,
  restored from NSSDCA data set PSPG-00559}\BibitemShut {NoStop}%
\bibitem [{\citenamefont {Grimm}(2018)}]{Grimm2018}%
  \BibitemOpen
  \bibfield  {author} {\bibinfo {author} {\bibfnamefont {R.~E.}\ \bibnamefont
  {Grimm}},\ }\bibfield  {title} {\bibinfo {title} {New analysis of the
  {Apollo} 17 surface electrical properties experiment},\ }\href
  {https://doi.org/10.1016/j.icarus.2018.06.007} {\bibfield  {journal}
  {\bibinfo  {journal} {Icarus}\ }\textbf {\bibinfo {volume} {314}},\ \bibinfo
  {pages} {389} (\bibinfo {year} {2018})}\BibitemShut {NoStop}%
\bibitem [{\citenamefont {Kong}(1972)}]{Kong1972}%
  \BibitemOpen
  \bibfield  {author} {\bibinfo {author} {\bibfnamefont {J.~A.}\ \bibnamefont
  {Kong}},\ }\bibfield  {title} {\bibinfo {title} {Electromagnetic fields due
  to dipole antennas over stratified anisotropic media},\ }\href
  {https://doi.org/10.1190/1.1440321} {\bibfield  {journal} {\bibinfo
  {journal} {Geophysics}\ }\textbf {\bibinfo {volume} {37}},\ \bibinfo {pages}
  {985} (\bibinfo {year} {1972})}\BibitemShut {NoStop}%
\bibitem [{\citenamefont {Tsang}\ \emph {et~al.}(1973)\citenamefont {Tsang},
  \citenamefont {Kong},\ and\ \citenamefont {Simmons}}]{Tsang1973}%
  \BibitemOpen
  \bibfield  {author} {\bibinfo {author} {\bibfnamefont {L.}~\bibnamefont
  {Tsang}}, \bibinfo {author} {\bibfnamefont {J.~A.}\ \bibnamefont {Kong}},\
  and\ \bibinfo {author} {\bibfnamefont {G.}~\bibnamefont {Simmons}},\
  }\bibfield  {title} {\bibinfo {title} {Interference patterns of a horizontal
  electric dipole over layered dielectric media},\ }\href
  {https://doi.org/10.1029/JB078i017p03287} {\bibfield  {journal} {\bibinfo
  {journal} {J. Geophys. Res.}\ }\textbf {\bibinfo {volume} {78}},\ \bibinfo
  {pages} {3287} (\bibinfo {year} {1973})}\BibitemShut {NoStop}%
\bibitem [{\citenamefont {Dasgupta}(2026{\natexlab{b}})}]{Dasgupta2026}%
  \BibitemOpen
  \bibfield  {author} {\bibinfo {author} {\bibfnamefont {P.}~\bibnamefont
  {Dasgupta}},\ }\href@noop {} {\bibinfo {title} {Exact spherical-wave forward
  model for radio reflection from stratified media and implications for the
  anomalous-polarity events observed by {ANITA}}} (\bibinfo {year}
  {2026}{\natexlab{b}}),\ \bibinfo {note} {submitted to Phys.\ Rev.\ D},\
  \Eprint {https://arxiv.org/abs/2608.15989} {arXiv:2608.15989 [astro-ph.IM]}
  \BibitemShut {NoStop}%
\bibitem [{\citenamefont {Sommerfeld}(1909)}]{Sommerfeld1909}%
  \BibitemOpen
  \bibfield  {author} {\bibinfo {author} {\bibfnamefont {A.}~\bibnamefont
  {Sommerfeld}},\ }\bibfield  {title} {\bibinfo {title} {{\"U}ber die
  {A}usbreitung der {W}ellen in der drahtlosen {T}elegraphie},\ }\href
  {https://doi.org/10.1002/andp.19093330402} {\bibfield  {journal} {\bibinfo
  {journal} {Ann. Phys.}\ }\textbf {\bibinfo {volume} {333}},\ \bibinfo {pages}
  {665} (\bibinfo {year} {1909})}\BibitemShut {NoStop}%
\bibitem [{\citenamefont {Weyl}(1919)}]{Weyl1919}%
  \BibitemOpen
  \bibfield  {author} {\bibinfo {author} {\bibfnamefont {H.}~\bibnamefont
  {Weyl}},\ }\bibfield  {title} {\bibinfo {title} {Ausbreitung
  elektromagnetischer {W}ellen {\"u}ber einem ebenen {L}eiter},\ }\href
  {https://doi.org/10.1002/andp.19193652104} {\bibfield  {journal} {\bibinfo
  {journal} {Ann. Phys.}\ }\textbf {\bibinfo {volume} {365}},\ \bibinfo {pages}
  {481} (\bibinfo {year} {1919})}\BibitemShut {NoStop}%
\bibitem [{\citenamefont {Chew}(1995)}]{Chew1995}%
  \BibitemOpen
  \bibfield  {author} {\bibinfo {author} {\bibfnamefont {W.~C.}\ \bibnamefont
  {Chew}},\ }\href {https://doi.org/10.1109/9780470547052} {\emph {\bibinfo
  {title} {Waves and Fields in Inhomogeneous Media}}}\ (\bibinfo  {publisher}
  {IEEE Press},\ \bibinfo {address} {New York},\ \bibinfo {year}
  {1995})\BibitemShut {NoStop}%
\bibitem [{\citenamefont {Parratt}(1954)}]{Parratt1954}%
  \BibitemOpen
  \bibfield  {author} {\bibinfo {author} {\bibfnamefont {L.~G.}\ \bibnamefont
  {Parratt}},\ }\bibfield  {title} {\bibinfo {title} {Surface studies of solids
  by total reflection of {X}-rays},\ }\href
  {https://doi.org/10.1103/PhysRev.95.359} {\bibfield  {journal} {\bibinfo
  {journal} {Phys. Rev.}\ }\textbf {\bibinfo {volume} {95}},\ \bibinfo {pages}
  {359} (\bibinfo {year} {1954})}\BibitemShut {NoStop}%
\bibitem [{\citenamefont {Werthm{\"u}ller}(2017)}]{Werthmuller2017}%
  \BibitemOpen
  \bibfield  {author} {\bibinfo {author} {\bibfnamefont {D.}~\bibnamefont
  {Werthm{\"u}ller}},\ }\bibfield  {title} {\bibinfo {title} {An open-source
  full {3D} electromagnetic modeler for {1D} {VTI} media in {P}ython:
  empymod},\ }\href {https://doi.org/10.1190/geo2016-0626.1} {\bibfield
  {journal} {\bibinfo  {journal} {Geophysics}\ }\textbf {\bibinfo {volume}
  {82}},\ \bibinfo {pages} {WB9} (\bibinfo {year} {2017})}\BibitemShut
  {NoStop}%
\bibitem [{\citenamefont {Geyer}(1992)}]{Geyer1992}%
  \BibitemOpen
  \bibfield  {author} {\bibinfo {author} {\bibfnamefont {C.~J.}\ \bibnamefont
  {Geyer}},\ }\bibfield  {title} {\bibinfo {title} {Practical {M}arkov chain
  {M}onte {C}arlo},\ }\href {https://doi.org/10.1214/ss/1177011137} {\bibfield
  {journal} {\bibinfo  {journal} {Stat. Sci.}\ }\textbf {\bibinfo {volume}
  {7}},\ \bibinfo {pages} {473} (\bibinfo {year} {1992})}\BibitemShut {NoStop}%
\bibitem [{\citenamefont {Olhoeft}\ and\ \citenamefont
  {Strangway}(1975)}]{Olhoeft1975}%
  \BibitemOpen
  \bibfield  {author} {\bibinfo {author} {\bibfnamefont {G.~R.}\ \bibnamefont
  {Olhoeft}}\ and\ \bibinfo {author} {\bibfnamefont {D.~W.}\ \bibnamefont
  {Strangway}},\ }\bibfield  {title} {\bibinfo {title} {Dielectric properties
  of the first 100 meters of the {M}oon},\ }\href
  {https://doi.org/10.1016/0012-821X(75)90146-6} {\bibfield  {journal}
  {\bibinfo  {journal} {Earth Planet. Sci. Lett.}\ }\textbf {\bibinfo {volume}
  {24}},\ \bibinfo {pages} {394} (\bibinfo {year} {1975})}\BibitemShut
  {NoStop}%
\bibitem [{\citenamefont {Robinson}\ and\ \citenamefont
  {Jolliff}(2002)}]{RobinsonJolliff2002}%
  \BibitemOpen
  \bibfield  {author} {\bibinfo {author} {\bibfnamefont {M.~S.}\ \bibnamefont
  {Robinson}}\ and\ \bibinfo {author} {\bibfnamefont {B.~L.}\ \bibnamefont
  {Jolliff}},\ }\bibfield  {title} {\bibinfo {title} {Apollo 17 landing site:
  {T}opography, photometric corrections, and heterogeneity of the surrounding
  highland massifs},\ }\href {https://doi.org/10.1029/2001JE001614} {\bibfield
  {journal} {\bibinfo  {journal} {J. Geophys. Res.}\ }\textbf {\bibinfo
  {volume} {107}},\ \bibinfo {pages} {5110} (\bibinfo {year}
  {2002})}\BibitemShut {NoStop}%
\bibitem [{\citenamefont {N\'evot}\ and\ \citenamefont
  {Croce}(1980)}]{NevotCroce1980}%
  \BibitemOpen
  \bibfield  {author} {\bibinfo {author} {\bibfnamefont {L.}~\bibnamefont
  {N\'evot}}\ and\ \bibinfo {author} {\bibfnamefont {P.}~\bibnamefont
  {Croce}},\ }\bibfield  {title} {\bibinfo {title} {Caract\'erisation des
  surfaces par r\'eflexion rasante de rayons {X}. {A}pplication \`a l'\'etude
  du polissage de quelques verres silicates},\ }\href
  {https://doi.org/10.1051/rphysap:01980001503076100} {\bibfield  {journal}
  {\bibinfo  {journal} {Rev. Phys. Appl.}\ }\textbf {\bibinfo {volume} {15}},\
  \bibinfo {pages} {761} (\bibinfo {year} {1980})}\BibitemShut {NoStop}%
\bibitem [{\citenamefont {Cooper}\ \emph {et~al.}(1974)\citenamefont {Cooper},
  \citenamefont {Kovach},\ and\ \citenamefont {Watkins}}]{Cooper1974}%
  \BibitemOpen
  \bibfield  {author} {\bibinfo {author} {\bibfnamefont {M.~R.}\ \bibnamefont
  {Cooper}}, \bibinfo {author} {\bibfnamefont {R.~L.}\ \bibnamefont {Kovach}},\
  and\ \bibinfo {author} {\bibfnamefont {J.~S.}\ \bibnamefont {Watkins}},\
  }\bibfield  {title} {\bibinfo {title} {Lunar near-surface structure},\ }\href
  {https://doi.org/10.1029/RG012i003p00291} {\bibfield  {journal} {\bibinfo
  {journal} {Rev. Geophys. Space Phys.}\ }\textbf {\bibinfo {volume} {12}},\
  \bibinfo {pages} {291} (\bibinfo {year} {1974})}\BibitemShut {NoStop}%
\bibitem [{\citenamefont {Bugiolacchi}\ \emph {et~al.}(2016)\citenamefont
  {Bugiolacchi}, \citenamefont {Bamford}, \citenamefont {Tar}, \citenamefont
  {Thacker}, \citenamefont {Crawford}, \citenamefont {Joy}, \citenamefont
  {Grindrod},\ and\ \citenamefont {Lintott}}]{Bugiolacchi2016}%
  \BibitemOpen
  \bibfield  {author} {\bibinfo {author} {\bibfnamefont {R.}~\bibnamefont
  {Bugiolacchi}}, \bibinfo {author} {\bibfnamefont {S.}~\bibnamefont
  {Bamford}}, \bibinfo {author} {\bibfnamefont {P.}~\bibnamefont {Tar}},
  \bibinfo {author} {\bibfnamefont {N.}~\bibnamefont {Thacker}}, \bibinfo
  {author} {\bibfnamefont {I.~A.}\ \bibnamefont {Crawford}}, \bibinfo {author}
  {\bibfnamefont {K.~H.}\ \bibnamefont {Joy}}, \bibinfo {author} {\bibfnamefont
  {P.~M.}\ \bibnamefont {Grindrod}},\ and\ \bibinfo {author} {\bibfnamefont
  {C.}~\bibnamefont {Lintott}},\ }\bibfield  {title} {\bibinfo {title} {The
  {Moon Zoo} citizen science project: {P}reliminary results for the {Apollo} 17
  landing site},\ }\href {https://doi.org/10.1016/j.icarus.2016.01.021}
  {\bibfield  {journal} {\bibinfo  {journal} {Icarus}\ }\textbf {\bibinfo
  {volume} {271}},\ \bibinfo {pages} {30} (\bibinfo {year} {2016})}\BibitemShut
  {NoStop}%
\bibitem [{\citenamefont {Heffels}\ \emph {et~al.}(2017)\citenamefont
  {Heffels}, \citenamefont {Knapmeyer}, \citenamefont {Oberst},\ and\
  \citenamefont {Haase}}]{Heffels2017}%
  \BibitemOpen
  \bibfield  {author} {\bibinfo {author} {\bibfnamefont {A.}~\bibnamefont
  {Heffels}}, \bibinfo {author} {\bibfnamefont {M.}~\bibnamefont {Knapmeyer}},
  \bibinfo {author} {\bibfnamefont {J.}~\bibnamefont {Oberst}},\ and\ \bibinfo
  {author} {\bibfnamefont {I.}~\bibnamefont {Haase}},\ }\bibfield  {title}
  {\bibinfo {title} {Re-evaluation of {Apollo} 17 {Lunar Seismic Profiling
  Experiment} data},\ }\href {https://doi.org/10.1016/j.pss.2016.11.007}
  {\bibfield  {journal} {\bibinfo  {journal} {Planet. Space Sci.}\ }\textbf
  {\bibinfo {volume} {135}},\ \bibinfo {pages} {43} (\bibinfo {year}
  {2017})}\BibitemShut {NoStop}%
\bibitem [{\citenamefont {Senior}\ and\ \citenamefont
  {Volakis}(1995)}]{SeniorVolakis1995}%
  \BibitemOpen
  \bibfield  {author} {\bibinfo {author} {\bibfnamefont {T.~B.~A.}\
  \bibnamefont {Senior}}\ and\ \bibinfo {author} {\bibfnamefont {J.~L.}\
  \bibnamefont {Volakis}},\ }\href {https://doi.org/10.1049/PBEW041E} {\emph
  {\bibinfo {title} {Approximate Boundary Conditions in Electromagnetics}}}\
  (\bibinfo  {publisher} {IEE},\ \bibinfo {address} {London},\ \bibinfo {year}
  {1995})\BibitemShut {NoStop}%
\bibitem [{\citenamefont {Li}\ \emph {et~al.}(2020)\citenamefont {Li},
  \citenamefont {Su}, \citenamefont {Pettinelli} \emph {et~al.}}]{Li2020}%
  \BibitemOpen
  \bibfield  {author} {\bibinfo {author} {\bibfnamefont {C.}~\bibnamefont
  {Li}}, \bibinfo {author} {\bibfnamefont {Y.}~\bibnamefont {Su}}, \bibinfo
  {author} {\bibfnamefont {E.}~\bibnamefont {Pettinelli}}, \emph {et~al.},\
  }\bibfield  {title} {\bibinfo {title} {The {M}oon's farside shallow
  subsurface structure unveiled by {Chang'E-4} {L}unar {P}enetrating {R}adar},\
  }\href {https://doi.org/10.1126/sciadv.aay6898} {\bibfield  {journal}
  {\bibinfo  {journal} {Sci. Adv.}\ }\textbf {\bibinfo {volume} {6}},\ \bibinfo
  {pages} {eaay6898} (\bibinfo {year} {2020})}\BibitemShut {NoStop}%
\bibitem [{\citenamefont {Zhang}\ \emph {et~al.}(2025)\citenamefont {Zhang},
  \citenamefont {Ding}, \citenamefont {Su}, \citenamefont {Shen}, \citenamefont
  {Lu}, \citenamefont {Soldovieri}, \citenamefont {Xiao}, \citenamefont {Zeng},
  \citenamefont {Du}, \citenamefont {Liu}, \citenamefont {Jiang}, \citenamefont
  {Gennarelli}, \citenamefont {Catapano}, \citenamefont {Dai}, \citenamefont
  {Feng}, \citenamefont {Wang}, \citenamefont {Liu},\ and\ \citenamefont
  {Li}}]{Zhang2025CE6}%
  \BibitemOpen
  \bibfield  {author} {\bibinfo {author} {\bibfnamefont {Z.}~\bibnamefont
  {Zhang}}, \bibinfo {author} {\bibfnamefont {C.}~\bibnamefont {Ding}},
  \bibinfo {author} {\bibfnamefont {Y.}~\bibnamefont {Su}}, \bibinfo {author}
  {\bibfnamefont {S.}~\bibnamefont {Shen}}, \bibinfo {author} {\bibfnamefont
  {W.}~\bibnamefont {Lu}}, \bibinfo {author} {\bibfnamefont {F.}~\bibnamefont
  {Soldovieri}}, \bibinfo {author} {\bibfnamefont {Z.}~\bibnamefont {Xiao}},
  \bibinfo {author} {\bibfnamefont {X.}~\bibnamefont {Zeng}}, \bibinfo {author}
  {\bibfnamefont {W.}~\bibnamefont {Du}}, \bibinfo {author} {\bibfnamefont
  {Y.}~\bibnamefont {Liu}}, \bibinfo {author} {\bibfnamefont {C.}~\bibnamefont
  {Jiang}}, \bibinfo {author} {\bibfnamefont {G.}~\bibnamefont {Gennarelli}},
  \bibinfo {author} {\bibfnamefont {I.}~\bibnamefont {Catapano}}, \bibinfo
  {author} {\bibfnamefont {S.}~\bibnamefont {Dai}}, \bibinfo {author}
  {\bibfnamefont {J.}~\bibnamefont {Feng}}, \bibinfo {author} {\bibfnamefont
  {Y.}~\bibnamefont {Wang}}, \bibinfo {author} {\bibfnamefont {J.}~\bibnamefont
  {Liu}},\ and\ \bibinfo {author} {\bibfnamefont {C.}~\bibnamefont {Li}},\
  }\bibfield  {title} {\bibinfo {title} {Subsoil structure at the {Chang'E-6}
  landing site revealed by in-situ {Lunar Regolith Penetrating Radar}},\ }\href
  {https://doi.org/10.1038/s43247-025-02631-4} {\bibfield  {journal} {\bibinfo
  {journal} {Commun. Earth Environ.}\ }\textbf {\bibinfo {volume} {6}},\
  \bibinfo {pages} {640} (\bibinfo {year} {2025})}\BibitemShut {NoStop}%
\bibitem [{\citenamefont {Bradley}\ \emph {et~al.}(2019)\citenamefont
  {Bradley}, \citenamefont {Tauscher}, \citenamefont {Rapetti},\ and\
  \citenamefont {Burns}}]{Bradley2019}%
  \BibitemOpen
  \bibfield  {author} {\bibinfo {author} {\bibfnamefont {R.~F.}\ \bibnamefont
  {Bradley}}, \bibinfo {author} {\bibfnamefont {K.}~\bibnamefont {Tauscher}},
  \bibinfo {author} {\bibfnamefont {D.}~\bibnamefont {Rapetti}},\ and\ \bibinfo
  {author} {\bibfnamefont {J.~O.}\ \bibnamefont {Burns}},\ }\bibfield  {title}
  {\bibinfo {title} {A ground plane artifact that induces an absorption profile
  in averaged spectra from global 21 cm measurements, with possible application
  to {EDGES}},\ }\href {https://doi.org/10.3847/1538-4357/ab0d8b} {\bibfield
  {journal} {\bibinfo  {journal} {Astrophys. J.}\ }\textbf {\bibinfo {volume}
  {874}},\ \bibinfo {pages} {153} (\bibinfo {year} {2019})}\BibitemShut
  {NoStop}%
\bibitem [{\citenamefont {Politis}\ and\ \citenamefont
  {Romano}(1992)}]{Politis1992}%
  \BibitemOpen
  \bibfield  {author} {\bibinfo {author} {\bibfnamefont {D.~N.}\ \bibnamefont
  {Politis}}\ and\ \bibinfo {author} {\bibfnamefont {J.~P.}\ \bibnamefont
  {Romano}},\ }\bibfield  {title} {\bibinfo {title} {A circular
  block-resampling procedure for stationary data},\ }in\ \href@noop {} {\emph
  {\bibinfo {booktitle} {Exploring the Limits of Bootstrap}}},\ \bibinfo
  {editor} {edited by\ \bibinfo {editor} {\bibfnamefont {R.}~\bibnamefont
  {LePage}}\ and\ \bibinfo {editor} {\bibfnamefont {L.}~\bibnamefont
  {Billard}}}\ (\bibinfo  {publisher} {Wiley},\ \bibinfo {address} {New York},\
  \bibinfo {year} {1992})\ pp.\ \bibinfo {pages} {263--270}\BibitemShut
  {NoStop}%
\bibitem [{\citenamefont {Politis}\ and\ \citenamefont
  {Romano}(1994)}]{PolitisRomano1994}%
  \BibitemOpen
  \bibfield  {author} {\bibinfo {author} {\bibfnamefont {D.~N.}\ \bibnamefont
  {Politis}}\ and\ \bibinfo {author} {\bibfnamefont {J.~P.}\ \bibnamefont
  {Romano}},\ }\bibfield  {title} {\bibinfo {title} {The stationary
  bootstrap},\ }\href {https://doi.org/10.1080/01621459.1994.10476870}
  {\bibfield  {journal} {\bibinfo  {journal} {J. Am. Stat. Assoc.}\ }\textbf
  {\bibinfo {volume} {89}},\ \bibinfo {pages} {1303} (\bibinfo {year}
  {1994})}\BibitemShut {NoStop}%
\end{thebibliography}%

\section*{End Matter}

\phantomsection\label{app:A}\textit{Appendix A: Field of the transmitter.}---Consider a horizontal electric dipole of moment $Il$ along $x$ at height $h_t$, and a receiver at height $h_r$, range $r$ and azimuth $\phi$ from the dipole axis ($\phi=90^\circ$ for the broadside and $0^\circ$ for the endfire transmitter). The coefficients $\mathcal R^{s,p}$ are $\mathcal R_0$ of Eq.~(\ref{eq:parratt}) for the two polarizations, with the Fresnel coefficients $r_{j-1,j}=(q_{j-1}-q_j)/(q_{j-1}+q_j)$ and the normalized admittances $q_j=k_{z,j}/k$ ($s$) and $k_{z,j}/(k\varepsilon_j)$ ($p$). With $\mathcal E=e^{ik_{z,0}(h_t+h_r)}$ and $J_n$ the Bessel function of the first kind of order $n$, the reflected magnetic field, a sum of propagating ($\kappa<k$) and evanescent ($\kappa>k$) waves, is \cite{DasguptaPRD}
\begin{align}
  H^{\rm ref}_x &= \frac{Il}{8\pi}\sin2\phi\int_0^\infty\!\kappa\,(\mathcal R^s+\mathcal R^p)\,J_2(\kappa r)\,\mathcal E\,d\kappa,\nonumber\\
  H^{\rm ref}_y &= -\frac{Il}{8\pi}\int_0^\infty\!\kappa\,\big[(\mathcal R^s-\mathcal R^p)J_0(\kappa r)\nonumber\\
        &\qquad+(\mathcal R^s+\mathcal R^p)\cos2\phi\,J_2(\kappa r)\big]\,\mathcal E\,d\kappa,\nonumber\\
  H^{\rm ref}_z &= \frac{iIl}{4\pi}\sin\phi\int_0^\infty\!\frac{\kappa^2}{k_{z,0}}\,\mathcal R^s\,J_1(\kappa r)\,\mathcal E\,d\kappa .
  \label{eq:sep}
\end{align}
Over a perfect conductor, $\mathcal R^s=-\mathcal R^p=-1$, and it reduces to the field of the image dipole. The receiver measures this field plus the direct field.

We obtain the field of the finite transmitter by integrating Eq.~(\ref{eq:sep}) over its current $I(x')=\cos(\pi x'/L_{\rm el})$, where $x'$ is measured from the center of a half-wave element of length $L_{\rm el}=\lambda/\{2[(1+3.3)/2]^{1/2}\}$, tuned for a surface permittivity of 3.3 \cite{Simmons1972}, with $h_t=5$~cm and $h_r=2.5$~m. We tabulate the integrals once for each frequency and description on a grid in range and sample the integrands densely near the branch point $\kappa=k$ and the guided-wave poles. Interpolation changes the fields by at most $2.2\times10^{-3}$ \cite{DasguptaPRD}. The continuous profiles are represented by 320 (graded), 240 (surrogate) and 64 (transition, over $z_t\pm8\zeta$) uniform layers, enough for convergence at these frequencies.

Beyond its far-field distance, at most 33~m, the transmitter acts as a point dipole. Replacing it by one changes the fields by 0.22~dB rms at 2.1~MHz and by at most 0.02~dB at 8.1--32.1~MHz. The fitted offsets absorb the amplitude of the current but not its distribution along the wire, which we test with the nine transmitter currents of the main text. As numerical checks, the recursion agrees with a direct solution of the boundary conditions at every interface and conserves power to $2\times10^{-14}$, and the integrals agree with their evaluation along a contour below the real axis to $2.3\times10^{-9}$ \cite{DasguptaPRD}.

\begin{table*}[t]
  \caption{Scores of the descriptions in each band, relative to the best description in that band. Zero marks the best, and a negative entry is the amount by which a description scores lower. The descriptions are fitted at 2.1 and 4.0~MHz, except the fixed surrogate, and at 8.1--32.1~MHz only their channel offsets are refitted. The OU columns give the log-likelihood of the covariance model of \hyperref[app:B]{Appendix~B}, evaluated for the best fit of each description and therefore not for the second maximum (dash). At 32.1~MHz only ranges below 1~km are used.\label{tab:fits}}
  \begin{ruledtabular}
  \setlength{\tabcolsep}{4pt}
  \begin{tabular}{llccccccc}
  & & & \multicolumn{2}{c}{2.1 + 4.0~MHz} & \multicolumn{2}{c}{8.1 + 16.0~MHz} & \multicolumn{2}{c}{32.1~MHz} \\
  & & & \multicolumn{2}{c}{fitted} & \multicolumn{2}{c}{predicted} & \multicolumn{2}{c}{predicted} \\
  Description & Parameters & $\tan\delta$ & $\leff$ & OU & $\leff$ & OU & $\leff$ & OU \\
  \hline
  two-layer & $\varepsilon'_1=4.27$, $\varepsilon'_2=6.03$, $d=17.9$~m & 0.008 & 0 & 0 & $-70.9$ & $-788.9$ & $-9.2$ & $-631.0$ \\
  graded & $\varepsilon'_0=2.18$, $\varepsilon'_\infty=6.01$, $z_g=11.3$~m & 0.001 & $-18.2$ & $-60.1$ & $-3.5$ & $-4.6$ & $-0.1$ & $-0.5$ \\
  two-layer, 2nd maximum & $\varepsilon'_1=2.89$, $\varepsilon'_2=5.06$, $d=14.9$~m & 0.018 & $-24.1$ & -- & $-28.8$ & -- & $-2.6$ & -- \\
  surrogate, fixed & $\varepsilon'_0=3.7$, $\varepsilon'_\infty=6.1$, $z_g=12$~m & 0.012 & $-36.5$ & $-84.6$ & 0 & 0 & $-0.1$ & 0 \\
  half-space & $\varepsilon'=1.68$ & 0.05 & $-169.0$ & $-111.2$ & $-5.7$ & $-23.3$ & 0 & $-0.4$ \\
  \end{tabular}
  \end{ruledtabular}
\end{table*}

\begin{table}[!htb]
  \caption{Rms fluctuation in dB about a smooth trend in the broadside channels, V the vertical field and H the horizontal power. The trend is a Gaussian smoothing with a standard deviation of 25~m, and the rms is taken over 100--800~m. The predictions use the fits at 2.1 and 4.0~MHz, with 2L the best two-layer fit and 2L$'$ its second maximum. The ratio is the 2L ripple divided by the measured fluctuation, so above 1 the predicted ripple exceeds it. Its spread covers trend widths of 25--60~m and the full traverse, 50--1500~m (50--1000~m at 32.1~MHz).\label{tab:ripple}}
  \begin{ruledtabular}
  \setlength{\tabcolsep}{2.6pt}
  \begin{tabular}{lccccccc}
  $f$ (MHz) & & Measured & 2L & Graded & 2L$'$ & Ratio & Spread \\
  \hline
  2.1 & V & 2.90 & 1.24 & 1.14 & 0.77 & 0.43 & 0.43--0.55 \\
  2.1 & H & 2.21 & 0.73 & 0.56 & 0.54 & 0.33 & 0.33--0.49 \\
  4.0 & V & 2.02 & 0.95 & 0.16 & 0.19 & 0.47 & 0.40--0.61 \\
  4.0 & H & 1.98 & 1.56 & 0.16 & 0.16 & 0.79 & 0.74--0.79 \\
  8.1 & V & 3.08 & 2.80 & 0.17 & 1.61 & 0.91 & 0.73--0.91 \\
  8.1 & H & 1.25 & 2.79 & 0.17 & 1.71 & 2.23 & 1.19--2.23 \\
  16.0 & V & 1.87 & 3.97 & 0.12 & 3.13 & 2.12 & 1.39--2.12 \\
  16.0 & H & 1.42 & 3.69 & 0.13 & 2.81 & 2.60 & 1.44--2.60 \\
  32.1 & V & 0.95 & 1.37 & 0.15 & 0.63 & 1.44 & 1.12--1.55 \\
  32.1 & H & 1.06 & 2.22 & 0.15 & 1.08 & 2.08 & 1.54--2.08 \\
  \end{tabular}
  \end{ruledtabular}
\end{table}

\begin{table}[!htb]
  \caption{Descriptions fitted jointly to 2.1--16~MHz, with their parameters below each row and scores relative to the best in each column. The sharp limit of the transition profile, $\zeta\le0.5$~m, coincides with the two-layer description \cite{DasguptaPRD}.\label{tab:joint}}
  \begin{ruledtabular}
  \setlength{\tabcolsep}{3pt}
  \begin{tabular}{lcccccc}
  & & \multicolumn{3}{c}{$\leff$ (bands in MHz)} & & \\
  Description & $\tan\delta$ & 2.1--4.0 & 8.1--16.0 & Sum & $z_5$ (m) & $\langle\varepsilon'\rangle$ \\
  \hline
  transition & 0.001 & 0 & $-17.2$ & 0 & 18.3 & 4.2 \\
  \multicolumn{7}{l}{\quad$\varepsilon'_{\rm top}=3.88$, $\varepsilon'_{\rm deep}=6.04$, $z_t=18.1$~m, $\zeta=3.3$~m} \\
  two-layer & 0.025 & $-21.1$ & 0 & $-3.8$ & 11.5 & 3.4 \\
  \multicolumn{7}{l}{\quad$\varepsilon'_1=3.37$, $\varepsilon'_2=5.92$, $d=11.5$~m} \\
  graded & 0.001 & $-11.6$ & $-20.9$ & $-15.2$ & 14.7 & 4.0 \\
  \multicolumn{7}{l}{\quad$\varepsilon'_0=2.55$, $\varepsilon'_\infty=6.02$, $z_g=12.0$~m} \\
  \end{tabular}
  \end{ruledtabular}
\end{table}

\phantomsection\label{app:B}\textit{Appendix B: Data and statistics.}---We bin the levels in range, in bins of 10, 5 and 2.5~m at 2.1--4.0, 8.1--16.0 and 32.1~MHz. At 32.1~MHz we leave out ranges beyond 1~km, where the curvature of the Moon can shift the phase of reflected waves by more than 0.2~rad. At 1.0~MHz the whole 70~m wire radiated, with a current distribution that is not documented, and the traverse spans only 0.2--5 wavelengths. The comparison at this frequency therefore depends on the assumed current and receiver height \cite{DasguptaPRD}. The graded profile still reproduces the interference minima at 1.0~MHz, though not at their measured positions, whereas the SEP experiment team had rejected smooth profiles because a 20-layer model of that kind produced no interference pattern \cite{SEPFinal1975}.

The offset of each channel maximizes Eq.~(\ref{eq:elleff}), censored terms included, and is unique because the channel score is concave in it. The autocorrelation time is $\tau_c=1+2\sum_{m\ge1}\gamma_{c,m}$, where $\gamma_{c,m}$ is the autocorrelation of the binned residuals at a lag of $m$ bins, with the sum truncated by the initial-positive-sequence rule \cite{Geyer1992}. We take its median over six reference descriptions, which is 2--27 bins at 2.1--4.0~MHz and 6--85 bins at 8.1--32.1~MHz.

In the covariance model, the residuals of each channel are a zero-mean Gaussian process with covariance $\sigma_{\rm OU}^2e^{-|r_i-r_{i'}|/r_{\rm OU}}$ between bins at ranges $r_i$ and $r_{i'}$. Its likelihood factorizes exactly along the traverse. We fit the amplitude $\sigma_{\rm OU}$ and the correlation length $r_{\rm OU}$ for each channel and description and omit the censored bins.

The bootstrap resamples each channel in blocks of $\max(3,\lceil\tau_c\rceil)$ bins, with five schemes. Circular blocks \cite{Politis1992} and blocks of geometrically distributed length \cite{PolitisRomano1994} are drawn in each channel separately. Common blocks, with full and half length, take the same stretches of range in every channel. Stratified circular blocks are drawn separately within 0--300, 300--800 and beyond 800~m. For the comparisons at 2.1--4.0~MHz we refit each replicate in a model linearized about the optimum, with the two-layer description refitted from both of its maxima.

The linearization fails where the fitted surface permittivity reaches its upper bound, in 16--19\% of the common-block replicates. In a set of 80 common-block replicates, we therefore refit in full all replicates at the bound and all others in which the two descriptions lie within 10 of each other. These full refits reduce the number of replicates in which the graded profile leads from 8 to 4, or 5\% (95\% Wilson interval 2--12\%). Away from the bound, they change the difference between the two descriptions by at most 8.7. The ripple statistic of Table~\ref{tab:ripple} and Fig.~\ref{fig:ripple}(a) uses the detected bins only.

\textit{Reverse test.}---Fitted at 8.1--16.0~MHz alone, both descriptions overestimate the deep permittivity, which those frequencies barely sample. Their predictions at 2.1--4.0~MHz then score 38.9 (graded) and 47.2 (two-layer) below their own fits there. In the forward direction, the graded profile fitted at 2.1--4.0~MHz predicts 8.1--16.0~MHz within 3.5 of its own fit there, whereas the two-layer description falls 92.3 below its own fit \cite{DasguptaPRD}.

\end{document}